\documentstyle[11pt,newpasp,twoside,epsf,amstex]{article}
\begin{document}
\title{RR\,Lyrae variables in globular clusters and nearby galaxies}
\author{M. Catelan}
\affil{Pontificia Universidad Cat\'olica de Chile, Departamento de Astronom\'\i a y Astrof\'\i sica, Av. Vicu\~{n}a Mackenna 4860, 782-0436 Santiago, Chile}

\begin{abstract}
I point out that the Oosterhoff dichotomy for globular cluster and field RR\,Lyrae stars may place the strongest constraints so far on the number of dwarf spheroidal-like protogalactic fragments that may have contributed to the formation of the Galactic halo. The first calibration of the RR\,Lyrae period-luminosity relation in $I$, $J$, $H$, $K$ taking evolutionary effects into account is provided. Problems in the interpretation of RR\,Lyrae light curves and evolutionary properties are briefly reviewed.
\end{abstract}

\section{Introduction}

Unmistakably old, with ages comparable to the age of the Universe, RR\,Lyrae (RRL) stars are an easily identified type of variable star which were clearly ``eyewitnesses'' of the formation of their parent galaxy. Therefore, they may provide precious information about the processes that led to the formation of galaxies in general, and of our own Milky Way and its system of satellite galaxies in particular. In this review, I discuss how the RRL star properties may constrain the possibility that the Galactic halo may have been built up from protogalactic fragments similar to the Galaxy's dwarf spheroidal (dSph) satellites. The RRL star period-luminosity (PL) relation is also presented, and open problems in the area are briefly discussed.

\section{RR\,Lyrae stars and the formation of the Galaxy}

Unavane, Wyse, \& Gilmore (1996) have attempted to place strong constraints on the number of dSph galaxies that may have contributed to the formation of the Galactic halo by studying its well-defined turnoff color. Finding very few stars bluer than the turnoff point in the halo field, but plenty in dSphs (see also Aparicio, these proceedings), they concluded that only a small fraction of the Galactic halo may have formed from the accretion of dSph-like protogalactic fragments.

However, this argument is limited in its scope: the young blue stars we see today were not present when the bulk of the halo formed. Accordingly, their argument only provides us with information about the relatively recent history of the halo. To put meaningful constraints on the extent to which the {\em bulk} of the halo may have formed by the accretion of ``protogalactic fragments'' or ``building blocks'' resembling the present-day dSphs (Searle \& Zinn 1978; Zinn 1993), we should really ``summon'' the right witnesses, i.e. those stars which saw it all happen, and survived to this day to tell us the whole truth about the matter.

RRL stars are ideal for this purpose. If the Galactic halo was built up from dSph-like fragments, the halo RRL stars must logically reflect the properties of the dSph RRL stars. In what follows, I shall try to provide hints on whether this is the case or not by tackling globular cluster (GC) and field RRL stars in turn.

\begin{figure}[t]
\plottwo{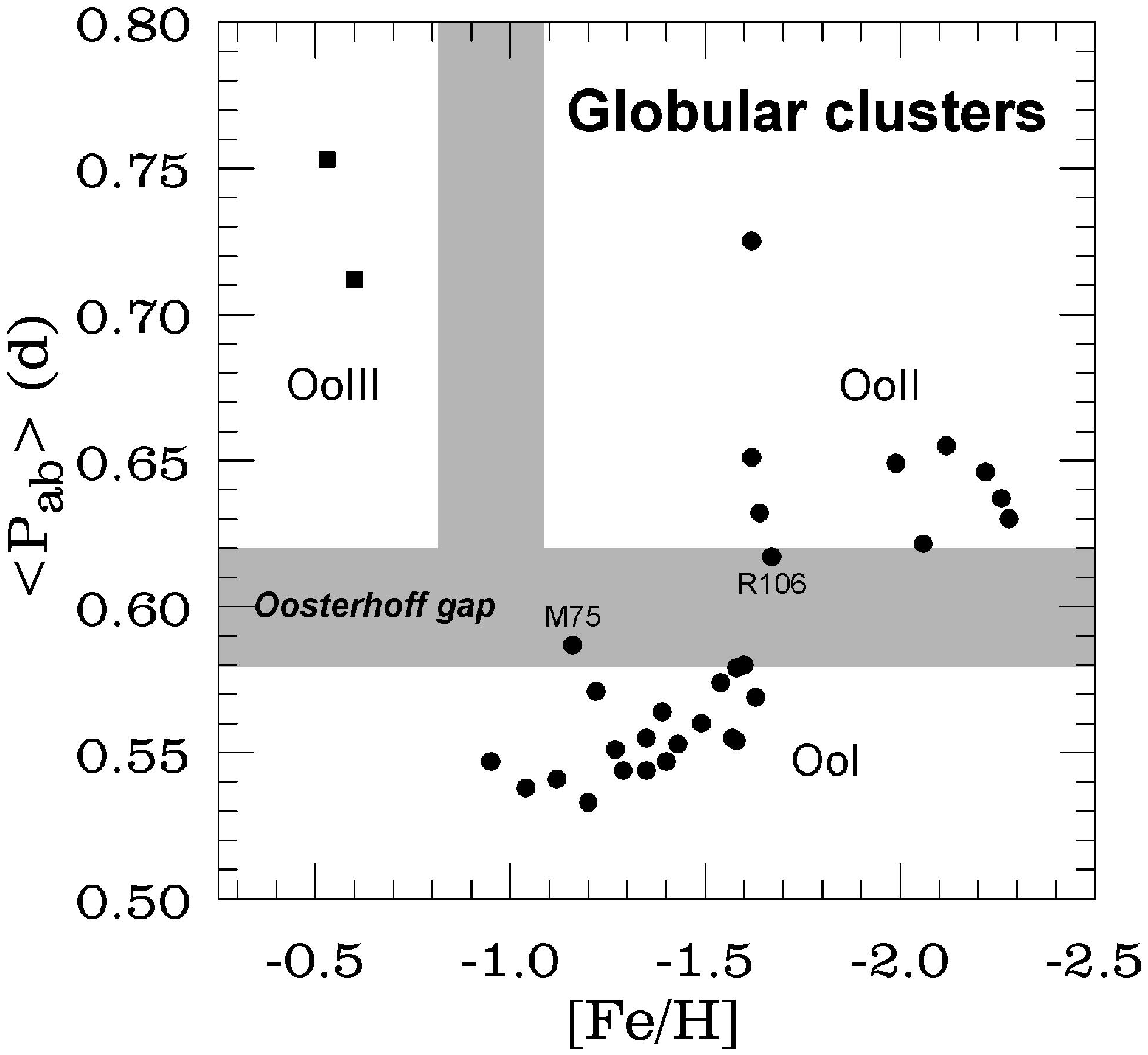}{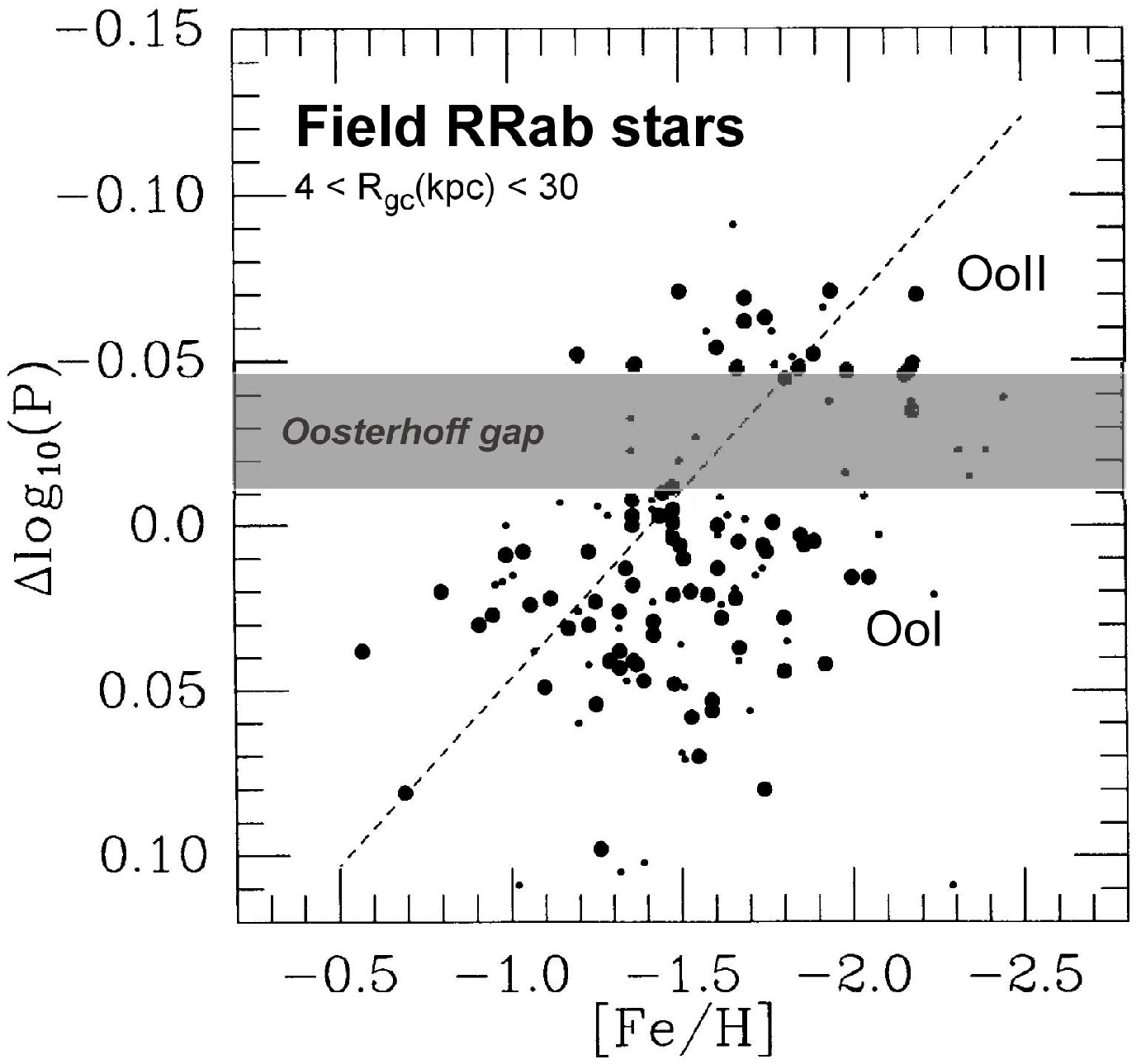}
\caption{Oosterhoff dichotomy in GCs ({left}) and in the field ({right}). $\langle P_{\rm ab}\rangle$ values were taken from Clement et al. (2001), except for M75 (Corwin et al. 2003), NGC\,6441 (Pritzl et al. 2003), and M68 ($\langle P_{\rm ab}\rangle$ recomputed from the Clement et al. online catalogue). The only GC with fewer than 10 known RRab stars in this plot is the OoIII GC NGC\,6388. In the field stars plot (adapted from SKK91), period shifts $\Delta$log$P$ were measured with respect to the M3 RRL stars, with the longer-period variables having {\em more negative} period shifts (see SKK91 for details). }
\end{figure}

One of the most interesting properties of RRL stars in Galactic GCs is the so-called {\em Oosterhoff dichotomy} (Oosterhoff 1939). In Fig.\,1 ({left}), I show the distribution of the mean periods of ab-type RRL stars, for GCs containing at least 10 RRL stars. As has long been known, Galactic GCs tend to clump around two main regions in this diagram, one with $0.52 \la \langle P_{\rm ab}({\rm d})\rangle \la 0.58$ and ${\rm [Fe/H]} \ga -1.65$ (OoI group), and the other with $0.62 \la \langle P_{\rm ab}({\rm d})\rangle \la 0.66$ and ${\rm [Fe/H]} \la -1.6$ (OoII group). The period range in between 0.58\,d and 0.62\,d, the ``Oosterhoff gap,'' is avoided by Galactic GCs.\footnote{ The two GCs that are found inside the ``gap,'' M75 and Ruprecht\,106, are both peculiar in their own right: M75 has a trimodal horizontal branch (HB) (Corwin et al. 2003), whereas Rup\,106 is a ``young'' red HB GC with $[\alpha/{\rm Fe}] \approx 0$. Rup\,106's RRL stars all clump around the instability strip red edge (Kaluzny, Krzeminski, \& Mazur 1995), so that its long $\langle P_{\rm ab}\rangle$ is clearly due to a peculiar distribution of RRL star temperatures. This is supported by the fact that {\em all} its 13 known RRL stars are RRab stars.}

As indicated, a third Oosterhoff group has recently been identified (Pritzl et al. 2000), with even longer $\langle P_{\rm ab}\rangle$ than OoII GCs, but even more metal-rich than OoI GCs. Due to the predominance of red HB stars in OoIII GCs, their specific frequency of RRL stars is probably systematically lower than in groups OoI or OoII. In this sense, possible OoIII GCs may also include Terzan\,5 (Edmonds et al. 2001), 47\,Tucanae (Carney, Storm, \& Williams 1993), and (less likely) NGC\,6304 (Valenti, Bellazzini, \& Cacciari 2003), with a single (long-$P$) RRab star each, which however all have consistently longer periods than field RRL stars with similar [Fe/H].

While the existence of the Oosterhoff dichotomy among Galactic GCs has been generally appreciated, the same cannot be said with respect to its occurrence among {\em field stars}, which was demonstrated by Suntzeff, Kinman, \& Kraft (1991, SKK91). Fig.\,1 ({right}) is based on SKK91's Fig.\,8b, clearly showing that the two Oosterhoff groups can be identified among {\em individual} field halo stars, over the galactocentric distance range 4\,$\la R_{\rm gc}$\,(kpc)\,$\la$\,30. (The smaller dots in this diagram refer to RRL stars with variable or ill-determined amplitudes.)

\begin{table}[t]
\caption{The Oosterhoff types of Milky Way dSph satellites}
\vspace{2mm}\centering
\begin{tabular}{cccc}\hline\noalign{\smallskip}
System & $\langle {\rm [Fe/H]} \rangle$ & $\langle P{\rm ab} \rangle$ & Oo type \\
\hline\noalign{\smallskip}
Ursa Minor & $-2.2$ & 0.638 & II \\
Draco & $-2.0$ & 0.615 & Int \\
Carina & $-2.0$ & 0.631 & Int \\
Leo II & $-1.9$ & 0.619 & Int \\
Sculptor & $-1.8$ & 0.587 & Int \\
Leo I & $-1.7$ & 0.602 & Int \\
Sextans & $-1.7$ & 0.606 & Int \\
Fornax & $-1.3$ & 0.595 & Int \\
Sagittarius & $-1.0$ & 0.574 & I-Int \\
\noalign{\smallskip}\hline\hline
\end{tabular}
\end{table}

Galaxies in the immediate vicinity of the Milky Way show a very different picture. LMC GCs are well known to preferentially {\em occupy} the Oosterhoff gap region shown in Fig.\,1 (Bono, Caputo, \& Stellingwerf 1994). This immediately rules out the possibility that any building blocks of the Milky Way may have resembled the LMC some 12\,Gyr ago (or else the Oosterhoff dichotomy would not exist). Mackey \& Gilmore (2003) have recently argued that the Fornax dSph GCs too are Oosterhoff-intermediate. Cacciari, Bellazzini, \& Colucci (2002) have found that M54 (in the Sagittarius dSph) is Oosterhoff-intermediate as well. (In terms of Fig.\,1, which does not use any of their data, M54 is one of the GCs barely intersecting the Oosterhoff-gap band at ${\rm [Fe/H]} \approx -1.6$.) Other Sagittarius dSph GCs may include Arp\,2, Terzan\,7, Terzan\,8 (Da Costa \& Armandroff 1995), Palomar\,12 (Dinescu et al. 2000), and NGC\,5634 (Bellazzini, Ferraro, \& Ibata 2002), none of which have a sufficient number of known RRL stars for an Oosterhoff type determination. An (unbound) star cluster has been suggested to be present in the Ursa Minor dSph (Kleyna et al. 2003), though I am personally unaware of any attempt to specifically study any RRL stars that may be associated with this structure.

The Oosterhoff status of dSphs is summarized in Table~1, which provides an update over Mateo (1996) and Pritzl et al. (2002a), based on work by Cseresnjes (2001), Clementini (these proceedings), Dall'Ora et al. (2003), Kinemuchi \& Smith (private communication), and Siegel \& Majewski (2000). Clearly, Galactic dSphs are predominantly Oosterhoff-intermediate (as is And\,VI in M31; Pritzl et al. 2002a). However, it is important that the Oosterhoff type of a dSph be firmly based on the properties of their individual RRL stars, using diagrams similar to Fig.\,1 ({right}), and not solely on $\langle P_{\rm ab}\rangle$ (Mateo, Fischer, \& Krzeminski 1995; Siegel \& Majewski 2000; Pritzl et al. 2002a). The reason why this is important is that a mix of OoI and OoII populations could easily lead to a $\langle P_{\rm ab} \rangle$ value in the Oosterhoff-intermediate range. For instance, taking all the Galactic GCs, one finds $\langle P_{\rm ab} \rangle = 0.585$\,d (Clement et al. 2001), which would make the Galaxy an Oosterhoff-intermediate entity -- which it obviously is not! Still, the current evidence, for the dSphs for which adequate data for individual RRL stars is available, does seem to support their Oosterhoff-intermediate classification.

This, then, would seem to leave little room for identifying the present-day dSph galaxies as Searle-Zinn ``building blocks.'' While we do see evidence of ongoing mergers between some dSphs and the Galaxy (e.g., Ibata, Gilmore, \& Irwin 1995; Majewski et al. 2000; Palma et al. 2003), the Oosterhoff argument suggests that these must not have provided a major contribution to the stellar content of the halo. A similar conclusion follows from a comparison between the detailed abundance ratios in metal-poor dSph and halo stars (e.g., Shetrone et al. 2003).

To close, we note that Vivas \& Zinn (2002) have recently claimed, based on preliminary {\sc quest} results, that the Oosterhoff dichotomy is {\em not} present among halo field stars, in sharp contrast with SKK91. The reason for the discrepancy between the two studies is unclear at present.

\begin{figure}
\plotone{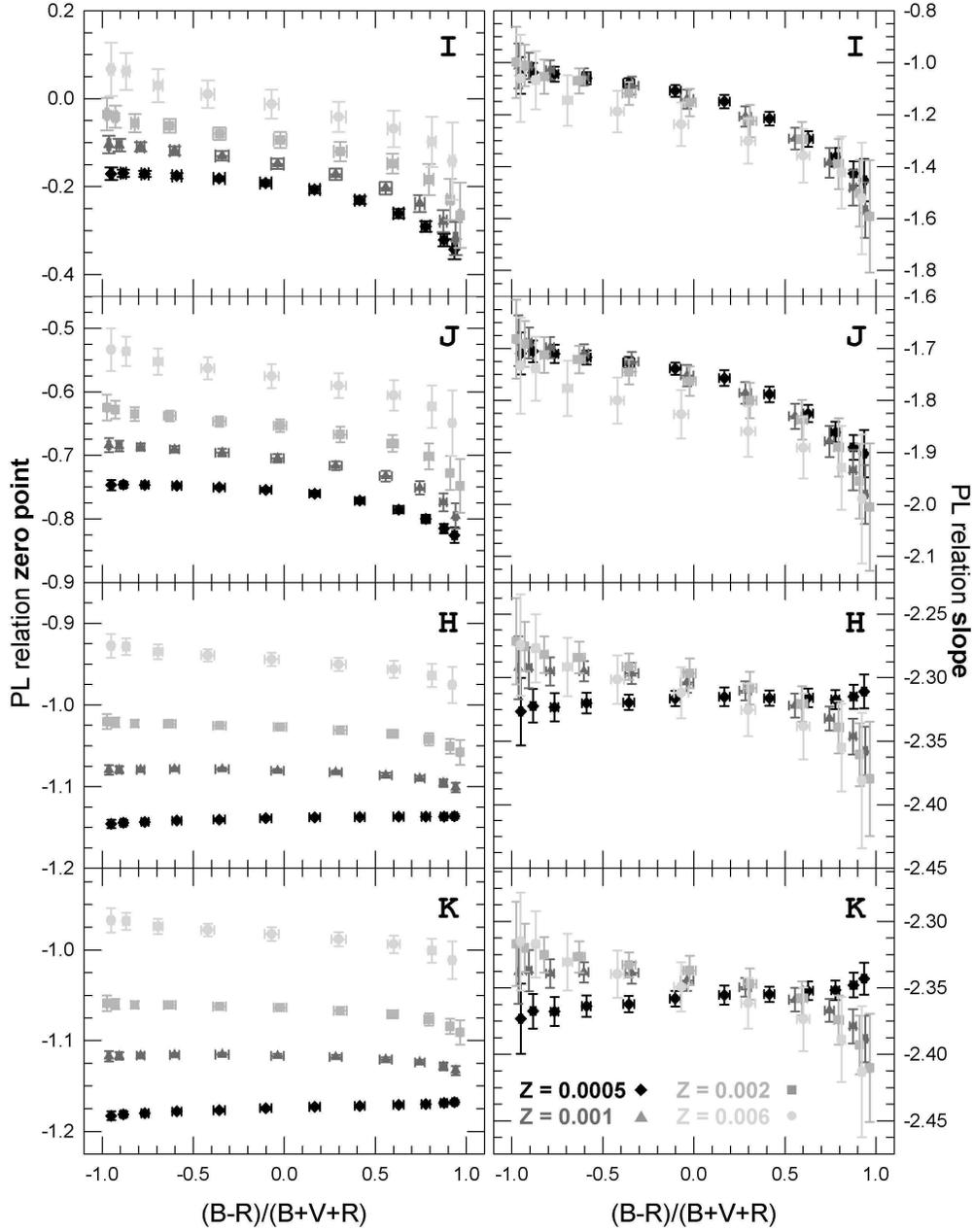}
\caption{PL relation for RRL stars. From top to bottom, for each of the filters $I$, $J$, $H$, $K$, the {left panels} show the zero points and the {right panels} show the slopes of the $M_X = f(\log P)$ relation as a function of the HB type. In each panel, different metallicities are indicated by different symbols and shades of gray.}
\end{figure}

\section{The RR\,Lyrae Period-Luminosity relation}

While the existence of an RRL PL relation in $K$ has long been recognized (Longmore, Fernley, \& Jameson 1986), a systematic analysis of RRL PL relations in other bandpasses has not been carried out to date. Using bandpasses in which the HB is not quite ``horizontal'' at the RRL level, one expects that a PL relation should be present. This is the case, in particular, in $I$, $J$, $H$, $K$. The predicted existence of a PL relation in $I$ is particularly interesting, in view of the wider availability and ease of $I$-band observations.

In Fig.\,2, I show theoretically calibrated PL relations in $I$, $J$, $H$, $K$, with individual rows showing the run of zero point ({left panels}) and slope ({right panels}) of the $M_X = f(\log P)$ relation for each of these filters as a function of HB type, as given by the Lee-Zinn parameter ($B$--$R$)/($B$+$V$+$R$). Synthetic HBs were used in order to take into account evolutionary effects. Each data point represents the average over 100 HB simulations with 500 stars each, slopes and zero points having been obtained from the Isobe et al. (1990) ``OLS bisector'' technique. The synthetic HBs were computed as in Catelan (2004), with the addition of bolometric corrections for $J$, $H$, $K$ from Girardi et al. (2002). As the plots show, all bands present some dependence on both metallicity and HB type, though some of the effects are more prominent in $I$ and $J$. For the $I$ band, the relations can be well fit by third-order polynomials, as follows:
\begin{displaymath}
M_I = a + b \, \log P, \,\,\,\, {\rm with} \,\,\,\,\,
a = \sum_{i = 0}^{3} a_{i} \left(\frac{B-R}{B+V+R}\right)^{i}, \,\,\,\,
b = \sum_{i = 0}^{3} b_{i} \left(\frac{B-R}{B+V+R}\right)^{i}.
\end{displaymath}

\noindent The $a_i$, $b_i$ coefficients are provided in Table~2. Full details, along with an analysis of systematic effects, will be provided elsewhere (Catelan, Pritzl, \& Smith, in preparation).

\begin{table}[t]
\caption{RRL PL relation in $I$: coefficients of the fits}
\vspace{2mm}
\centering
\begin{tabular}{ccccc}
\hline\noalign{\smallskip}
Zero Point & $a_{0}$ & $a_{1}$ & $a_{2}$ &$a_{3}$ {\smallskip}\\
\hline\noalign{\smallskip}
$Z = 0.0005$ & $-0.19407$ & $-0.05497$ & $-0.06715$ & $-0.04099$ \\
$Z = 0.0010$ & $-0.14511$ & $-0.05621$ & $-0.06682$ & $-0.05938$ \\
$Z = 0.0020$ & $-0.08994$ & $-0.04445$ & $-0.06105$ & $-0.07562$ \\
$Z = 0.0060$ & $-0.01482$ & $-0.04683$ & $-0.02630$ & $-0.07607$ \\
\hline\noalign{\smallskip}
Slope & $b_{0}$ & $b_{1}$ & $b_{2}$ & $b_{3}$ {\smallskip} \\
\hline\noalign{\smallskip}
$Z = 0.0005$ & $-1.11959$ & $-0.17183$ & $-0.14633$ & $-0.05973$ \\
$Z = 0.0010$ & $-1.13554$ & $-0.16489$ & $-0.15966$ & $-0.13782$ \\
$Z = 0.0020 $ & $-1.14989$ & $-0.10576$ & $-0.13972$ & $-0.20024$  \\
$Z = 0.0060 $ & $-1.24590$ & $-0.11549$ & $-0.04421$ & $-0.14658$  \\
\noalign{\smallskip}\hline\hline
\end{tabular}
\end{table}

The comparison between these model predictions and the observations is limited by the lack of extensive datasets for GCs in the redder passbands. However, in Fig.\,3 I show the PL relations in $I$ for a metal-poor (M92, data from Kopacki 2001) and a metal-intermediate GC (IC\,4499, data from Walker \& Nemec 1996). The RRc and candidate RRe (i.e., second-overtone) stars were ``fundamentalized'' in the same way, by adding 0.128 to their log-periods. The linear regressions were carried out using the Isobe et al. (1990) OLS bisector method. In the case of IC\,4499, two RRc outliers (one of which is off scale) were not considered in the fits. There is good agreement between the empirical slopes and the theoretical calibration in Fig.\,2. Note also that the candidate RRe stars fall nicely along the locus defined by the RRab and RRcd stars in these plots, even though the procedure to fundamentalize their periods is strictly applicable only to the RRc stars. This may point to the possibility that they may simply be the short-period tail of the RRc class.

\section{Fourier decomposition woes?}

Fourier decomposition of RRL light curves may directly provide some of their physical parameters. In particular, calibrations for $M$ and $L$ based on hydrodynamic models for first-overtone pulsators (Simon \& Clement 1993) have been widely employed in the recent literature. These read as follows:

\vskip 0.15cm
\centerline{$\log L = 1.04 \, \log P - 0.058 \,\, \phi_{31} + 2.41$,}

\centerline{$\log M = 0.52 \, \log P - 0.11 \, \phi_{31} + ~0.39$,}
\vskip 0.15cm

\noindent where $\phi_{31} = \phi_3 - 3 \phi_1$. Combining the two equations and rearranging, one finds:

\vskip 0.15cm
\centerline{$\log P = -2.877 + 1.305\,\log L - 0.688\,\log M$.}
\vskip 0.15cm

\begin{figure}[t]
\plotone{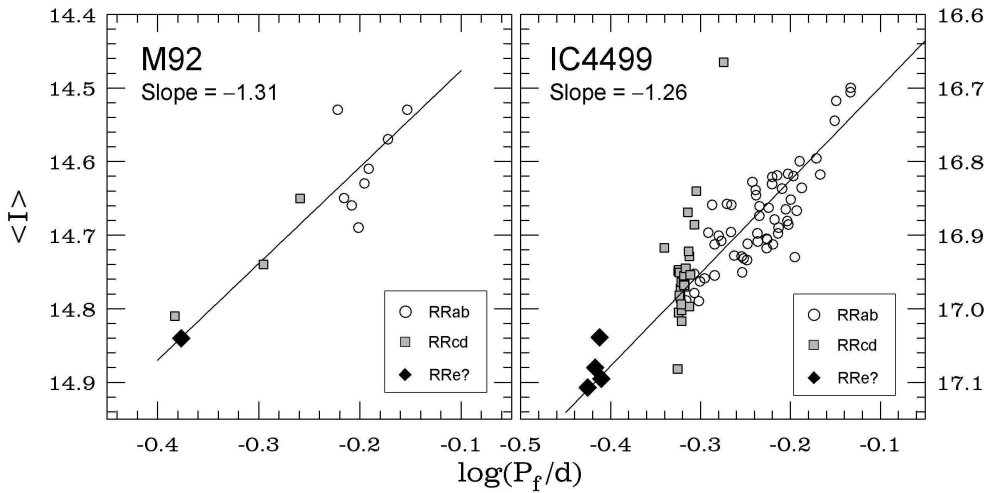}
\caption{Observed PL relation in $I$ for M92 ({left}) and for IC\,4499 ({right}). RRc and candidate RRe variables have been fundamentalized. }
\end{figure}

\noindent Clearly, there is a serious problem with this relation: it lacks a temperature term, which we know, from the period--mean density relation, should be present. The solution to this conundrum is unclear, but I suggest that the above relations for $L$ and $M$ cannot (both) be (simultaneously) valid.

\section{Epilogue: some important open problems}

Important open problems in the study of RRL stars that should be kept in mind when analyzing the RRL data for GCs and nearby galaxies include the following:

\begin{itemize}
\item How universal are local calibrations of the $M_V^{\rm RR} - {\rm [Fe/H]}$ relation? Since we know, through Ritter's relation, that the $M_V^{\rm RR} - {\rm [Fe/H]}$ relation is reflected upon the periods of RRL stars, one should make sure to check that the same $\langle P\rangle-{\rm [Fe/H]}$ progression (e.g., Fig.\,1) as in the calibrating sample is present in populations to which the relation is applied, which may easily not be the case if the underlying HB type--[Fe/H] trend is different. In addition, are ``second parameters'' at play in other galaxies (having, for instance, a different chemical enrichment history, or formed from faster rotating clouds of gas) that may directly change the HB luminosity with respect to the calibrating samples?

\item What causes the ``peaked'' period distribution in M3 and other GCs (Rood \& Crocker 1989; Catelan 2004)? Does HB evolution ``slow down'' when RRL stars are about to change pulsation mode? Does the Blazhko effect play a role?

\item What is the evolutionary status of RRL stars in OoII GCs? We still seem unable to account for the observed number of RRL stars in these GCs if they are evolved away from a position on the blue zero-age HB (Pritzl et al. 2002b).
\vspace{2mm}

\item Why do OoIII GC RRL stars have such dramatically long periods for their parent cluster metallicities, in sharp contrast with field RRL stars of similar [Fe/H]?
\vspace{2mm}

\item Why have $\omega$\,Cen, NGC\,6388, and NGC\,6441 succeeded in producing several long-period RRc stars ($P_{\rm c} > 0.45$\,d), while similar stars are entirely lacking in other GCs (with the exception of V70 in M3)?\footnote{ The intriguingly large number of long-period RRc stars in Fornax dSph GCs 1 and 3 is likely to be an artifact of the poor phase coverage in the Mackey \& Gilmore (2003) HST study. Support for this conclusion is provided by the fact that long-period RRc stars were not found in more extensive ground-based variability surveys (Maio et al. 2003; Clementini, private communication).} The fact that M3--V70 is peculiarly bright (Corwin \& Carney 2001) suggests that the above three clusters all have peculiarly bright HBs, which is also consistent with the location of long-period RRc stars in the theoretical period-amplitude diagrams (Bono et al. 1997).

\end{itemize}

\acknowledgments I thank H.A. Smith, B.J. Pritzl, G. Clementini, A.V. Sweigart, J. Borissova, T.M. Corwin, N.B. Suntzeff, and R.T. Rood for their comments and for interesting discussions over the course of the past several years. This work was supported by Proyecto FONDECYT Regular No. 1030954.

\section*{Discussion}
\noindent {\it Dambis:} Does the period-luminosity relation depend on population in the same way as the metallicity-$M_V$ relation?\\

\noindent {\it Catelan:} If you compare the results that I showed for IC\,4499, NGC\,6441, M68, and M92, you are led to conclude that, from an empirical perspective, the population dependence is relatively small. The detailed effect depends on the bandpass though.\\

\noindent {\it Comment by G. Bono:} The $K$-band PL relation when compared with the $I$-band PL relation presents 3 main advantages: 1) smaller dependence on reddening corrections; 2) smaller dependence on evolutionary effects; 3) smaller dependence on metal content.\\

\noindent {\it Catelan:} Indeed, I did not mean to imply that the PL relation in $K$ is not good; rather, I wanted to emphasize that we do not necessarily need to go to $K$ to have a PL relation which is also good. It is often the case that $I$-band data are more readily available to the observers, which I encourage to be used with our proposed PL relation in $I$. Another disadvantage of the $I$-band relation, compared to the $K$-band relation, is the former's stronger dependence on reddening effects.\\

\noindent {\it McNamara:} In NGC\,6388 I find that the RR\,Lyrae stars are fainter than RR\,Lyrae stars in clusters that contain Oosterhoff\,II RR\,Lyrae stars, contrary to your conclusions.\\

\noindent {\it Catelan:} I believe that your result is based on a single SX\,Phoenicis star that we found in the cluster, so that small-number statistics may be at play. Also, given the long mean periods that we find for this cluster's RR\,Lyrae stars, low luminosities would imply unrealistically low masses. Last, but not least, the Fourier decomposition parameters for RRab stars give luminosities comparable to those in OoII systems.\\

\noindent {\it Jurcsik:} What would you say is the cause of the Oosterhoff dichotomy? \\

\noindent {\it Catelan:} Let me first say what the traditional answer to this question has been. It is believed that there is a region of the instability strip where the RR\,Lyrae stars may pulsate either in the fundamental or in the first overtone mode -- the so-called ``either-or'' zone. Back in 1973, van Albada \& Baker suggested that the actual pulsation mode, in this area, is defined by a star's evolutionary history. Hence it was suggested that in OoII globulars, where HB stars may cross the instability strip while on their final redward excursion to the AGB, the ``either-or'' region will be populated by RRc stars. In OoI globulars, in turn, blueward-evolving RR\,Lyrae stars will also be present, at an earlier (i.e., slower) evolutionary stage -- so that the ``either-or'' zone will be primarily populated by RRab stars. The lack of globulars in between the two groups, as suggested by Castellani and by Renzini in the early-80s, would simply be due to the fact that, in the relevant metallicity range, Galactic globulars somehow develop exclusively blue HBs, without RR\,Lyrae variables being present in significant numbers. There are a few problems with this global scenario: i) A large overlap in color between RRc and RRab pulsators is expected in OoI globulars, but none is observed, at least in M3; ii) It is difficult to produce redward-evolving RR\,Lyrae stars in significant numbers in OoII GCs; iii) Sharp peaks in the period distributions, not predicted by the evolutionary or pulsation models, may be responsible, at least in part, for the Oosterhoff types. Until we are able to explain these open problems, I would say we do not completely understand the Oosterhoff dichotomy.\\

\noindent {\it Cassisi:} Since the evolutionary rates of HB models strongly depend on the efficiency of mixing processes in the convective core, have you performed any tests in order to verify the effects of different assumptions for mixing efficiency on period distributions predicted by HB models?\\

\noindent {\it Catelan:} No, I have not, but I strongly encourage people computing HB evolutionary tracks to perform such tests. This may be particularly important in connection with the failure of canonical models to produce enough RR\,Lyrae stars evolved away from a position on the blue ZAHB for OoII globular clusters such as M68 and M15.\\

%

\end{document}